\documentclass{pasj01}
\jyear{2022}
\Received{$\langle$reception date$\rangle$}
\Accepted{$\langle$acception date$\rangle$}
\Published{$\langle$publication date$\rangle$}
\usepackage{url}
\usepackage[T1]{fontenc}
\usepackage{ae,aecompl}

\begin{document}

\newcommand{\ssparea}{1310}
\newcommand{\sspfullarea}{984}
\newcommand{\Ngal}{103191}
\newcommand{\Ncand}{2319}
\newcommand{\NA}{8}
\newcommand{\NB}{28}
\newcommand{\NC}{138}

\title{Survey of Gravitationally-lensed Objects in HSC Imaging (SuGOHI). VIII. New galaxy-scale lenses from the HSC SSP}

\author{Kenneth C. \textsc{Wong}\altaffilmark{1}}
\email{kcwong19@gmail.com}

\author{James H. H. \textsc{Chan}\altaffilmark{2}}

\author{Dani C.-Y. \textsc{Chao}\altaffilmark{3,4,5}}

\author{Anton T. \textsc{Jaelani}\altaffilmark{6}}

\author{Issha \textsc{Kayo}\altaffilmark{7}}

\author{Chien-Hsiu \textsc{Lee}\altaffilmark{8}}

\author{Anupreeta \textsc{More}\altaffilmark{9,10}}

\author{Masamune \textsc{Oguri}\altaffilmark{11,12,10}}

\altaffiltext{1}{National Astronomical Observatory of Japan, 2-21-1 Osawa, Mitaka, Tokyo 181-8588, Japan}
\altaffiltext{2}{Institute of Physics, Laboratory of Astrophysics, Ecole Polytechnique F\'{e}d\'{e}rale de Lausanne (EPFL), Observatoire de Sauverny, 1290 Versoix, Switzerland}
\altaffiltext{3}{Ru{\dj}er Bo{\v s}kovi\'c Institute, Bijeni{\v c}ka cesta 54, 10000 Zagreb, Croatia}
\altaffiltext{4}{Max Planck Institute for Astrophysics, Karl-Schwarzschild-Str. 1, 85741 Garching, Germany}
\altaffiltext{5}{Physik-Department, Technische Universit{\"a}t M{\"u}nchen, James-Franck-Stra{\ss}e 1, 85748 Garching, Germany}
\altaffiltext{6}{Astronomy Research Group and Bosscha Observatory, FMIPA, Institut Teknologi Bandung, Jl. Ganesha 10, Bandung 40132, Indonesia}
\altaffiltext{7}{Department of Liberal Arts, Tokyo University of Technology, Ota, Tokyo 144-8650, Japan}
\altaffiltext{8}{NSF's National Optical-Infrared Astronomy Research Laboratory, 950 N. Cherry Avenue, Tucson, AZ 85719, USA}
\altaffiltext{9}{The Inter-University Centre for Astronomy and Astrophysics (IUCAA), Post Bag 4, Ganeshkhind, Pune 411007, India}
\altaffiltext{10}{Kavli IPMU (WPI), UTIAS, The University of Tokyo, Kashiwa, Chiba 277-8583, Japan}
\altaffiltext{11}{Center for Frontier Science, Chiba University, 1-33 Yayoi-cho, Inage-ku, Chiba 263-8522, Japan}
\altaffiltext{12}{Research Center for the Early Universe, University of Tokyo, Tokyo 113-0033, Japan}


\maketitle

\begin{abstract}
We conduct a search for galaxy-scale strong gravitational lens systems in Data Release 4 of the Hyper Suprime-Cam Subaru Strategic Program (HSC SSP), consisting of data taken up to the S21A semester.  We select \Ngal~luminous red galaxies from the Baryon Oscillation Spectroscopic Survey (BOSS) sample that have deep multiband imaging from the HSC SSP and use the {\sc YattaLens} algorithm to automatically identify lens candidates with blue arc-like features.  The candidates are visually inspected and graded based on their likelihood of being a lens.  We find \NA~definite lenses, \NB~probable lenses, and \NC~possible lenses.  The new lens candidates generally have lens redshifts in the range $0.3 \lesssim z_{\mathrm{L}} \lesssim 0.9$, a key intermediate redshift range to study the evolution of galaxy structure.  Follow-up spectroscopy will confirm these new lenses and measure source redshifts to enable detailed lens modeling.
\end{abstract}

\section{Introduction} \label{sec:intro}
Strong gravitational lensing by galaxies is a valuable tool for studying cosmology and extragalactic astronomy.  Since lensing is sensitive to the total mass distribution of the galaxy, including both baryonic and dark matter, it serves as a unique probe of galaxies' internal structure.  Strong lensing has been used to study the total density profile (e.g., \cite{koopmans+2006,koopmans+2009,auger+2010}), the stellar initial mass function (e.g., \cite{treu+2010,sonnenfeld+2019}), and the distribution of dark matter subhalos around the lens galaxy (e.g., \cite{vegetti+2010,vegetti+2012,vegetti+2014,nierenberg+2014,nierenberg+2017,gilman+2020}), among other properties.  Lensed quasars can also be used to constrain cosmological parameters, particularly the Hubble constant \citep{refsdal1964}, through a measurement of their time delays, providing an important independent measure of the expansion rate of the Universe (e.g., \cite{wong+2020,birrer+2020}).

However, strong lensing is a relatively rare phenomenon, requiring a chance alignment between a foreground massive galaxy or galaxy cluster with a bright background source.  Because of their rarity, deep and wide imaging surveys are required to build statistical samples of strong lenses.  In particular, lenses at intermediate redshifts ($0.3 \lesssim z_{\mathrm{L}} \lesssim 1.0$) are important for studying the evolution of galaxies over cosmic time.

The recently-completed Hyper Suprime Cam Subaru Strategic Program (HSC SSP; \cite{aihara+2018}) is an ideal dataset for discovering new gravitational lenses from deep multiband imaging data across a wide area of the sky.  The HSC SSP covers a large area ($> 1000$ deg$^{2}$) in the $grizy$ bands to a depth ($i\sim26.2$) greater than comparable surveys (e.g., Dark Energy Survey).  Our ongoing lens search, the Survey of Gravitationally lensed objects in HSC Imaging (SuGOHI), has already resulted in hundreds of new gravitational lens candidates at both the galaxy scale (SuGOHI-g; \cite{sonnenfeld+2018,wong+2018,sonnenfeld+2020}) and the galaxy group/cluster scale (SuGOHI-c; \cite{jaelani+2020}), including a number of lensed quasar candidates (SuGOHI-q; \cite{chan+2020}).  These lenses have been found through a variety of search methodologies, including automated algorithms, visual inspection, citizen science, and machine learning methods.

In this paper, we apply the {\sc YattaLens} algorithm \citep{sonnenfeld+2018} to the latest data release of the HSC SSP, covering \ssparea~deg$^{2}$ of the sky.  {\sc YattaLens} is a proven method that has been used in the past on HSC SSP data to discover new galaxy-scale lenses (e.g., \cite{sonnenfeld+2018,wong+2018}).  Starting with a catalog of known massive galaxies with spectroscopic redshifts, we use {\sc YattaLens} to automatically identify potential lens candidates in the multiband imaging data, which are then visually inspected and evaluated to find the most promising systems.

This paper is organized as follows.  In Section~\ref{sec:data}, we describe the imaging and spectroscopic data used in this study.  Section~\ref{sec:search} describes our lens search methodology.  In Section~\ref{sec:candidates}, we present our newly-discovered strong lens candidates.  We summarize our results and conclusions in Section~\ref{sec:summary}.  All magnitudes given are on the AB system.  All images are oriented with North up and East to the left.

\section{Data} \label{sec:data}

\subsection{HSC SSP Imaging} \label{subsec:hscssp}
The HSC SSP is an optical imaging survey conducted with the Hyper Suprime-Cam \citep{miyazaki+2012,miyazaki+2018,kawanomoto+2018,komiyama+2018} on the Subaru telescope, which has a pixel scale of $0\farcs168$/pixel.  The Wide component of the HSC SSP consists of $grizy$ imaging to a depth of $z\sim26.2$.  The data used in this paper are taken from Data Release 4 of the HSC SSP, which comprises data taken up to the S21A semester, and covers \ssparea~deg$^2$ in all bands, including \sspfullarea~deg$^2$ to the full depth.  The median seeing in the $i$-band is $\sim0\farcs6$.  The data are reduced with \texttt{hscPipe} version 8.4 \citep{bosch+2018}.

\subsection{BOSS Spectroscopy} \label{subsec:boss}
We pre-select objects to apply our search methodology to using spectroscopic data from the Baryon Oscillation Spectroscopic Survey (BOSS; \cite{dawson+2013}), a component of the Sloan Digital Sky Survey III (SDSS-III; \cite{eisenstein+2011}).  The BOSS sample consists primarily of luminous red galaxies (LRGs) out to a redshift of $z \lesssim 0.7$.  The redshifts used in this study come from Data Release 15 \citep{aguado+2019}.  In principle, this search could be scaled up to any input catalog of objects, with or without spectroscopic redshifts.  We use the BOSS LRG sample because these are the most massive galaxies, and therefore are more likely to be strong lenses due to their large lensing cross-section, as well as the fact that they have existing spectroscopy that could be valuable for follow-up.

\section{Lens Search} \label{sec:search}

\subsection{Pre-selection} \label{subsec:preselection}
We select galaxies from the HSC SSP that have spectroscopic redshifts from BOSS between $0.1 < z < 1.0$ and that have at least one observation in each of the $g$, $r$, and $i$ bands.  We do not include galaxies that were originally searched in \citet{sonnenfeld+2018}, unless the galaxy only had one $g$-band observation at the time, as a deeper observation could reveal lensing features.  After selecting the sample of galaxies and matching with the BOSS catalog, we remove known lenses by matching with the Master Lens Database (Moustakas et al. in preparation).  We also match against the existing SuGOHI catalog, which includes all previous candidates from previous SuGOHI publications, and remove these objects.  This leaves a total of \Ngal~galaxies to be processed.

\subsection{YattaLens} \label{subsec:yattalens}
We use the lens-finding algorithm {\sc YattaLens} \citep{sonnenfeld+2018} to select lens candidates from our input catalog.  Here, we briefly describe the {\sc YattaLens} algorithm and the settings we use to run it.

We start by obtaining cutouts roughly $10\arcsec \times 10\arcsec$ ($\sim$ $60 \times 60$ pixels at the HSC pixel scale) in size centered on each of the input objects in the $gri$ bands, as well as the corresponding point spread functions (PSF).  {\sc YattaLens} fits an elliptical de Vaucouleurs profile \citep{devaucouleurs1948} to the $i$-band image within a 3\arcsec~radius region around the center of the galaxy.  The best-fit profile is then convolved with the PSF and subtracted from all three bands.  {\sc YattaLens} then calls {\sc SExtractor} \citep{bertinarnouts1996} to search the lens-subtracted $g$-band cutout for objects that could be lensed arcs using the following criteria:
\begin{enumerate}
\item Distance from the lens center between 3 and 30 pixels (roughly $0\farcs5 < R < 5\farcs0$).
\item Major-to-minor axis ratio $>$ 1.4.
\item Maximum difference of $30^{\circ}$ between the position angle of the major axis of the object and the direction tangential to a circle centered on the lens and passing through the object centroid.
\item Minimum angular aperture (the angle subtended by the object as measured from the lens centroid) of $25^{\circ}$.
\item A footprint size between 20 and 500 pixels.
\end{enumerate}
Objects that satisfy all of these criteria are considered to be potential lensed arcs.  If at least one candidate arc is detected, the objects detected in the $g$-band are masked out in the corresponding $i$-band image and the main galaxy is fit again, this time with a more flexible S\'{e}rsic profile.  Additional non-arc-like objects in the $i$-band image are either fit with a S\'{e}rsic profile if it is within 1.3 times the distance of the farthest arc candidate from the lens, or are masked out of it is further away.

{\sc YattaLens} then applies a color cut of $g-i < 2.0$ to the lensed arc candidates.  This cut removes satellite galaxies that happen to be elongated tangential to the lens galaxy, which is a frequent source of contamination.  The object fluxes in the $g$ and $i$-bands are measured by summing over the pixels in the $g$-band footprint from {\sc SExtractor}, factoring in an additional 20\% systematic uncertainty that accounts for residuals in the lens light subtraction.  Different objects are considered to be multiple lensed images of the same source if their $g-i$ color is consistent within $2\sigma$.  Other arc-like objects with a different color that are within 1.3 times the distance of the farthest arc candidate from the lens are classified as foreground objects and are fitted with a S\'{e}rsic profile.  Further-away objects are masked out.  This process is then repeated for non-arc-like objects.

Each set of arcs is then fit with a simple lens model consisting of a singular isothermal ellipsoid (SIE).  The source light is modeled as a circular exponential profile.  The mass centroid is fixed to the lens light centroid.  The model's free parameters are the lens Einstein radius, lens axis ratio and associated position angle, the source position, source effective radius, and the amplitude of the lens, source, and foreground objects' surface brightness distributions.  Foreground objects are treated as massless and do not contribute to the lens potential.  The maximum-likelihood model fit is found using a modified version of the code developed by \citet{auger+2011}.  We do not report the lens model parameters, as YattaLens is designed purely for identifying lens candidates rather than detailed modeling, and the best-fit parameters can often be substantially off from the true values (A. Sonnenfeld, private communication).

{\sc YattaLens} also fits the image with a ring galaxy model and a S\'{e}rsic model with disky/boxy isophotes to see if either can fit the data better than the lens model.  This filters out common interlopers such as spiral arms and elongated foreground galaxies.  If the lens model has a higher $\chi^{2}$ than the ring or S\'{e}rsic models for each set of multiple image candidates, the object is deemed to not be a lens, otherwise it is considered to be a legitimate lens candidate.  We refer the reader to \citet{sonnenfeld+2018}, particularly Figures 1 and 2, for visual examples of this procedure.

After running {\sc YattaLens} on the \Ngal~objects in the input catalog, we are left with \Ncand~lens candidates.

\subsection{Lens Grading} \label{subsec:grading}
We visually inspect the lens candidates selected by {\sc YattaLens} and assign scores to them based on their probability of being a lens, $P_{\mathrm{lens}}$.  We score them on a numerical scale from 0 to 3 according to the following criteria:

\begin{itemize}
\item 3: definite lens ($P_{\mathrm{lens}} > 0.997$)
\item 2: probable lens ($0.997 > P_{\mathrm{lens}} > 0.5$)
\item 1: possible lens ($0.5 > P_{\mathrm{lens}} > 0.003$)
\item 0: non-lens ($0.003 > P_{\mathrm{lens}}$)
\end{itemize}

The scoring was performed by the eight co-authors of this study, all of whom have experience working with strong lensing images from the HSC SSP and have graded candidates in the past.  The $\chi^{2}$ information from {\sc YattaLens} is not used as part of the scoring procedure.  The \Ncand~candidates were initially split into two subsamples, which were each scored independently by four graders.  After this initial round of scoring, candidates that were unanimously agreed to be non-lenses were removed, and the remaining candidates were then scored by all eight graders.  The scores were then compiled and averaged, and the scores given by each grader were revealed and compared.  Graders were then allowed to discuss individual candidates and adjust their scores, particularly for controversial objects where the scores were discrepant or were close to the boundary of a lens grade.  The final classification of candidates is based on their mean score according to the following scheme, which has been used in previous SuGOHI publications:

\begin{itemize}
\item A (definite lens): $\langle$score$\rangle > 2.5$
\item B (probable lens): $2.5 \geq \langle$score$\rangle > 1.5$
\item C (possible lens): $1.5 \geq \langle$score$\rangle > 0.5$
\item D (non-lens): $\langle$score$\rangle \leq 0.5$
\end{itemize}

\section{New Lens Candidates} \label{sec:candidates}
\begin{figure*}
\includegraphics[width=\textwidth]{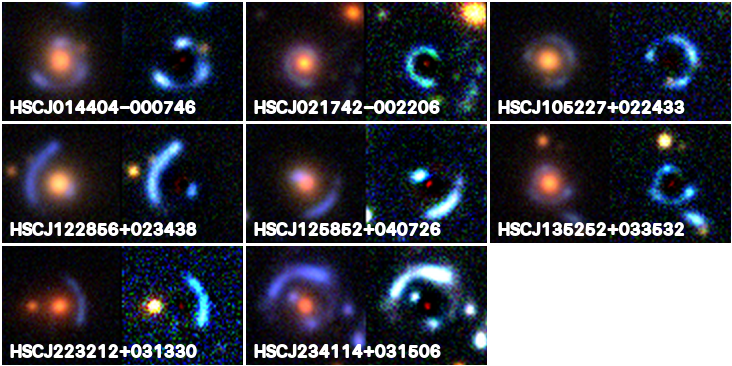}
\caption{Grade A lens candidates.  For each candidate, the left panel shows the composite $gri$ color image from the HSC SSP imaging, while the right panel shows the image with the lens galaxy subtracted by {\sc YattaLens}.  The cutouts are roughly $\sim 10\arcsec \times 10\arcsec$.}
\label{fig:gradeA}
\end{figure*}

In total, we discovered \NA~grade A candidates, \NB~grade B candidates, and \NC~grade C candidates.  The grade A and grade B candidates are listed in Table~\ref{tab:lens_candidates}.  We show multiband composite images of the grade A candidates in Figure~\ref{fig:gradeA} and of the grade B candidates in Figure~\ref{fig:gradeB}.  The lens redshift distribution is shown in Figure~\ref{fig:new_zlens_hist}.  The new lens candidates have spectroscopic lens redshifts spanning the range $0.3 \leq z_{\mathrm{L}} \leq 0.9$.  For comparison, we show the redshift distribution of our 36 grade A and grade B lenses in Figure~\ref{fig:ab_zlens_hist} along with the 106 previously-identified galaxy-galaxy lens candidates with grades A or B from the SuGOHI-g sample \citep{sonnenfeld+2018,sonnenfeld+2020,wong+2018,jaelani+2020} that have known lens redshifts.  For completeness, we list the grade C lens candidates in the Appendix.

We search the NASA/IPAC Extragalactic Database\footnote{http://ned.ipac.caltech.edu/} (NED) to see if spectroscopic source redshifts have been found for any of our grade A and B candidates.  We also check recent lens searches (Section~\ref{subsec:overlap}) for spectroscopic source redshift measurements.  These are indicated in Table~\ref{tab:lens_candidates} where available.  Interestingly, the three cases in which a source redshift is known are all grade B.  While the knowledge of a source redshift in these cases should, in principle, serve as strong evidence for the lensing nature of these objects (enough to elevate them to grade A), we report the original grade for consistency.  This, along with the somewhat lower proportion of grade A/B/C lenses relative to \citet{sonnenfeld+2018} and \citet{wong+2018}, does suggest that our grading is conservative in that the likelihood of a candidate being a lens may be higher than their grade suggests.

\renewcommand*\arraystretch{1.0}
\begin{table*}
\caption{Grade A and B Lens candidates \label{tab:lens_candidates}}
\begin{minipage}{\linewidth}
\begin{tabular}{l|rrrrcc}
\hline
Name &
$\alpha$ (J2000) &
$\delta$ (J2000) &
$z_{\mathrm{L}}$ &
$z_{\mathrm{S}}$ &
Grade &
Other refs.
\\
\hline
HSCJ001424+004145& 
    3.6012& 
    0.6959& 
0.5704& 
-- & 
B& 
1,9,11
\\
HSCJ012018+001125& 
   20.0756& 
    0.1905& 
0.5988& 
-- & 
B& 
9,10,11
\\
HSCJ014404--000746& 
   26.0168& 
   -0.1296& 
0.5631& 
-- & 
A& 
1
\\
HSCJ021742--002206& 
   34.4251& 
   -0.3685& 
0.6069& 
-- & 
A& 

\\
HSCJ092800+044653& 
  142.0036& 
    4.7815& 
0.7392& 
-- & 
B& 

\\
HSCJ093658+035845& 
  144.2438& 
    3.9792& 
0.5900& 
-- & 
B& 

\\
HSCJ104223+001521& 
  160.5974& 
    0.2559& 
0.5484& 
-- & 
B& 
5
\\
HSCJ104259+025858& 
  160.7459& 
    2.9830& 
0.5413& 
-- & 
B& 
2
\\
HSCJ104545+024149& 
  161.4409& 
    2.6971& 
0.5685& 
-- & 
B& 

\\
HSCJ104818+003434& 
  162.0766& 
    0.5762& 
0.5604& 
-- & 
B& 

\\
HSCJ105227+022433& 
  163.1130& 
    2.4094& 
0.5086& 
-- & 
A& 

\\
HSCJ111356+010404& 
  168.4865& 
    1.0680& 
0.6402& 
-- & 
B& 

\\
HSCJ111420+015149& 
  168.5852& 
    1.8639& 
0.5112& 
-- & 
B& 

\\
HSCJ115349+023128& 
  178.4559& 
    2.5247& 
0.7142& 
-- & 
B& 

\\
HSCJ115851+042737& 
  179.7126& 
    4.4604& 
0.5476& 
-- & 
B& 

\\
HSCJ120949+020908& 
  182.4542& 
    2.1525& 
0.7092& 
-- & 
B& 

\\
HSCJ122856+023438& 
  187.2369& 
    2.5772& 
0.4950& 
-- & 
A& 
2
\\
HSCJ123616+021421& 
  189.0668& 
    2.2394& 
0.4497& 
-- & 
B& 

\\
HSCJ123754+015902& 
  189.4752& 
    1.9839& 
0.6039& 
2.30\footnote{\cite{cao+2020}}& 
B& 
3
\\
HSCJ124125+042131& 
  190.3549& 
    4.3589& 
0.5027& 
-- & 
B& 

\\
HSCJ125516+041553& 
  193.8204& 
    4.2649& 
0.6548& 
-- & 
B& 

\\
HSCJ125852+040726& 
  194.7185& 
    4.1241& 
0.6046& 
-- & 
A& 

\\
HSCJ125925--001053& 
  194.8580& 
   -0.1814& 
0.6643& 
-- & 
B& 

\\
HSCJ130004+014450& 
  195.0202& 
    1.7473& 
0.5341& 
-- & 
B& 
6
\\
HSCJ130848+024602& 
  197.2003& 
    2.7675& 
0.5550& 
1.2551\footnote{\cite{talbot+2021}}& 
B& 
4
\\
HSCJ131513+003445& 
  198.8072& 
    0.5794& 
0.5989& 
-- & 
B& 
2
\\
HSCJ131945+013151& 
  199.9412& 
    1.5311& 
0.6971& 
-- & 
B& 
7
\\
HSCJ133350+022145& 
  203.4600& 
    2.3625& 
0.5686& 
-- & 
B& 
2
\\
HSCJ133459+001617& 
  203.7478& 
    0.2715& 
0.4836& 
-- & 
B& 
2
\\
HSCJ135252+033532& 
  208.2177& 
    3.5924& 
0.6302& 
-- & 
A& 

\\
HSCJ142822+031759& 
  217.0936& 
    3.3000& 
0.5885& 
1.3072\footnote{\cite{drinkwater+2017}}& 
B& 
7
\\
HSCJ150408+015744& 
  226.0373& 
    1.9624& 
0.5246& 
-- & 
B& 

\\
HSCJ223212+031330& 
  338.0507& 
    3.2252& 
0.7042& 
-- & 
A& 
9,11
\\
HSCJ234114+031506& 
  355.3116& 
    3.2518& 
0.7268& 
-- & 
A& 
8
\\
HSCJ234117--002937& 
  355.3230& 
   -0.4939& 
0.5314& 
-- & 
B& 
9,10,11
\\
HSCJ235853+012406& 
  359.7217& 
    1.4018& 
0.4805& 
-- & 
B& 
7,9,11
\\
\hline
\end{tabular}
\\
{\footnotesize Lens redshifts are from SDSS DR15.  ``Other refs." column indicates recent publications that also identified these systems as lens candidates, and are indicated as follows: 1) \citet{jacobs+2019}; 2) \citet{petrillo+2019}; 3) \citet{cao+2020}; 4) \citet{talbot+2021}; 5) \citet{li+2020}; 6) \citet{li+2021}; 7) \citet{huang+2021}; 8) \citet{stein+2022}; 9) \citet{canameras+2021}; 10) Jaelani et al. (in preparation); 11) \citet{shu+2022}}
\end{minipage}
\end{table*}
\renewcommand*\arraystretch{1.0}

\begin{figure*}
\includegraphics[width=\textwidth]{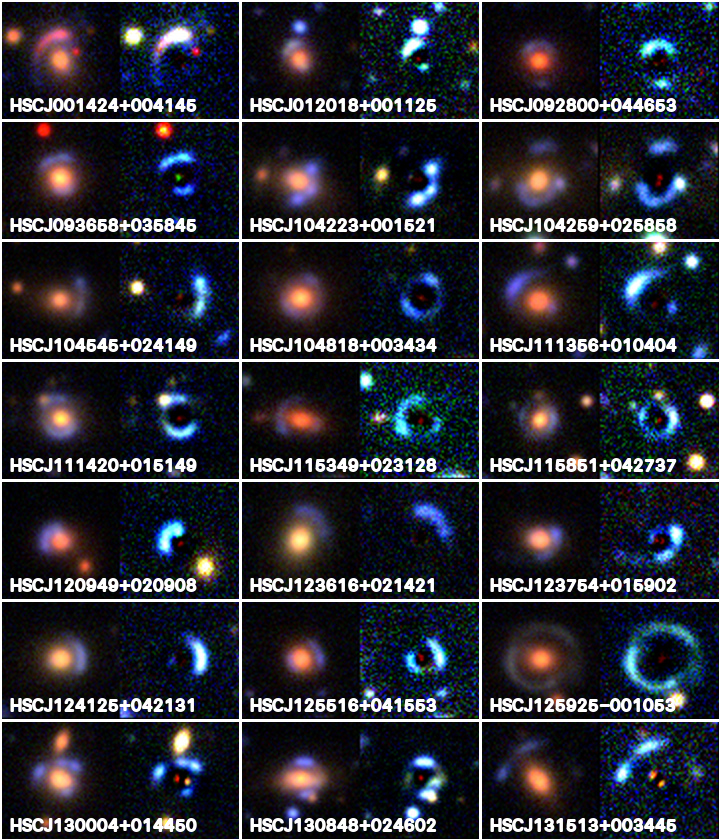}
\caption{Grade B lens candidates, similar to Figure~\ref{fig:gradeA}.}
\label{fig:gradeB}
\end{figure*}

\addtocounter{figure}{-1}
\begin{figure*}
\includegraphics[width=\textwidth]{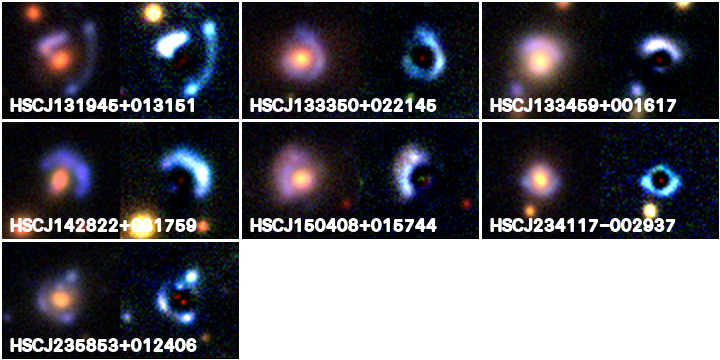}
\caption{(Continued.)}
\end{figure*}

\begin{figure}
\includegraphics[width=0.5\textwidth]{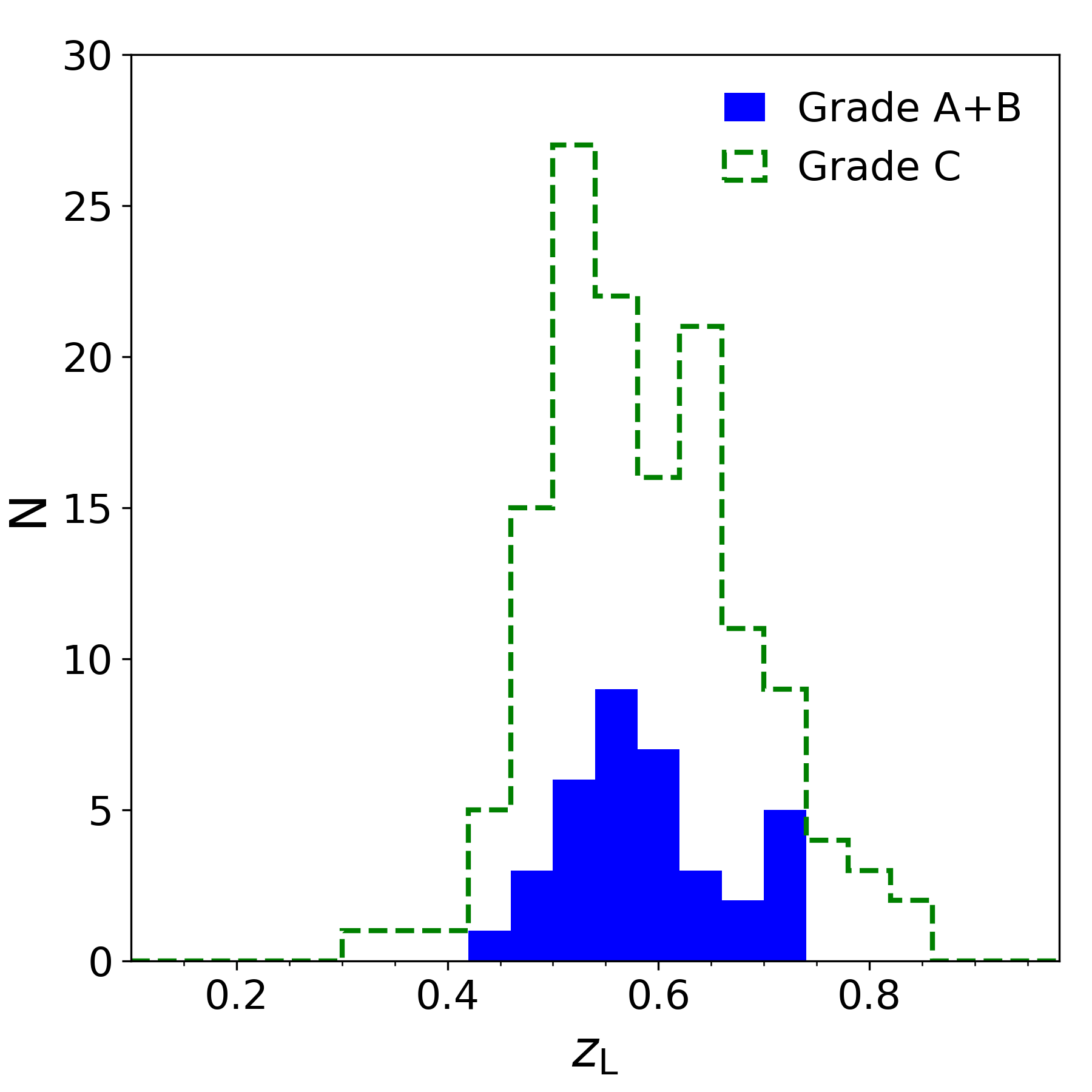}
\caption{Histogram of spectroscopic lens redshifts for the lens candidates found in this study.  The grade A and grade B candidates are combined (blue solid histogram) and shown separately from the grade C candidates (green dashed histogram).}
\label{fig:new_zlens_hist}
\end{figure}

\begin{figure}
\includegraphics[width=0.5\textwidth]{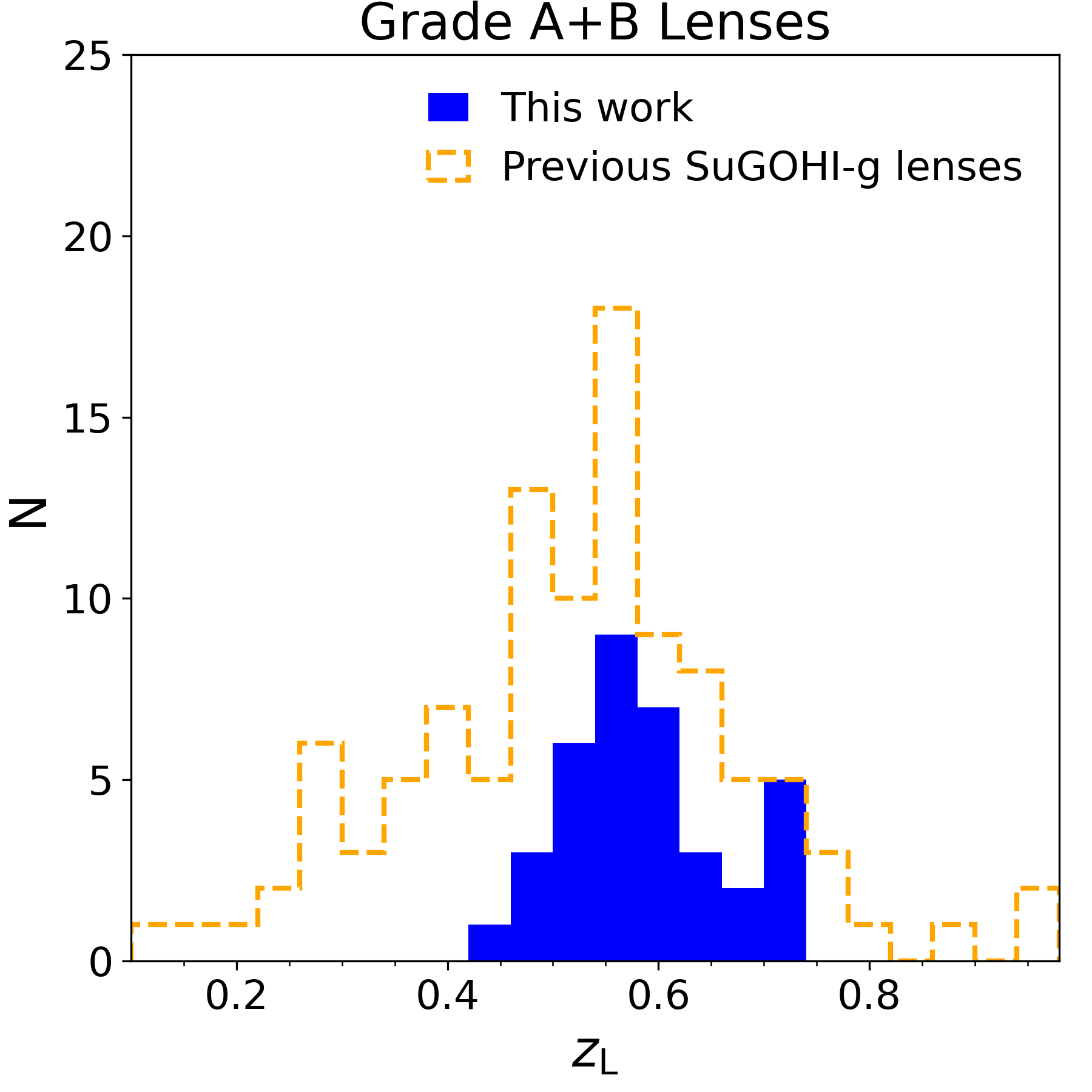}
\caption{Histogram of spectroscopic lens redshifts for the 36 combined grade A and grade B lens candidates found in this study (blue solid histogram), compared to lens redshifts for the existing 106 grade A and grade B galaxy-galaxy lenses in the SuGOHI-g sample (orange dashed histogram) that have known lens redshifts.}
\label{fig:ab_zlens_hist}
\end{figure}

\subsection{Overlap with other lens searches} \label{subsec:overlap}
Separate from this work, \citet{canameras+2021}, \citet{shu+2022}, and Jaelani et al. (in preparation) have recently applied machine learning-based lens search algorithms to the HSC SSP Public Data Release 2 \citep{aihara+2019} and have published lens candidates from these searches.  Additional lens candidates from recent lens searches in other surveys have also been compiled and are cross-matched with our candidates (R. Ca\~{n}ameras, private communication).

Although our work is based on a larger dataset, there is some overlap in the candidates we have discovered, and the quality of the HSC imaging data is often superior to those of these previous searches.  We have indicated these candidates and cited the appropriate references in Table~\ref{tab:lens_candidates}.  While some of these previous searches graded the lens candidates in a similar way to this study, they are often based on lower-quality data or used a slightly different grading scheme.  As such, we do not take into consideration a candidate's grade or likelihood of being a lens from these previous studies.

\subsection{Comparison to previous searches and datasets} \label{subsec:comparison}
We compare the statistics of lens candidates selected in this paper to previous searches in the HSC SSP using {\sc YattaLens}.  The \citet{sonnenfeld+2018} sample includes candidates selected through other methods, so it's not quite a fair comparison to this work.  \citet{wong+2018} used {\sc YattaLens} to analyze $\sim31000$ LRGs and identify 772 candidates, while this paper analyzes \Ngal~LRGs and returns \Ncand~candidates, which is a comparable rate.  After inspection, \citet{wong+2018} finds 200 definite/probable/possible candidates out of 776, while this paper finds 174 such candidates out of \Ncand.  This difference can be partially explained by our pre-selection, which removed $\sim150$ known and previously-selected candidates from entire SuGOHI sample of galaxy-scale lenses.  Had these systems not been filtered out, many of them likely would have been identified by {\sc YattaLens} and graded as possible candidates.  The lower rate in this paper may also be partially due to more conservative scoring, as was suggested in Section~\ref{subsec:overlap}.

It is difficult to directly compare the number lens candidates found in various surveys due to different survey depths, image quality, and area, as well as differences in the criteria used to judge the quality of a lens candidate.  Nevertheless, we can comment on a few comparable datasets and the size of the of lens candidate samples from them to the SuGOHI-g sample to give a broad overview of where our sample stands in relation to them.  Restricting the sample to galaxy-galaxy lenses with a grade of either A, B, or C from SuGOHI publications (i.e., not including external publications such as \cite{canameras+2021}, \cite{shu+2022}, etc.), we have identified 1464 lens candidates (of which 282 are grade A or B) to date in the HSC SSP in \ssparea~deg$^{2}$ through the latest data release.  By comparison, a search for strong lenses\footnote{https://kids.strw.leidenuniv.nl/DR4/hqlenses.php} in the Kilo-Degree Survey found 268 ``high-quality" candidates \citep{petrillo+2019,li+2020,li+2021} in $\sim1350$ deg$^{2}$.
A search through $\sim30000$ deg$^{2}$ in Pan-STARRS by \citet{canameras+2020} found 353 candidates with a grade roughly equivalent to our grade A and B lenses.  Finally, the survey most comparable to the HSC SSP is the Dark Energy Survey, in which lens searches through $\sim5000$ deg$^{2}$ have yielded $\sim500$ ``probable" lenses and $\sim750$ ``possible" lenses \citep{jacobs+2019}, with an additional 186 candidates identified by \citet{rojas+2021}.  In this context, the HSC SSP is quite competitive in number of lens candidates identified relative to survey area.  It should be noted again that the greater depth of the HSC SSP makes us more likely to detect fainter and higher-redshift lenses, which fills a valuable niche.

\subsection{Comments on Individual Lens Candidates}  \label{subsec:indiv}
Here, we briefly comment on a few notable systems among the grade A and B candidates that we have discovered.

\subsubsection{Group-scale lenses} \label{subsubsec:grouplens}
Some of our lens candidates are potentially group-scale lenses, as they either have multiple galaxies within the arc(s), or have an arc radius that implies a much larger lensing mass than a single lens galaxy.  These systems are HSCJ104545+024149 (grade B), HSCJ131945+013151 (grade B), and HSCJ223212+031330 (grade A).  These candidates, if confirmed, may be valuable for studying group-scale halos, particularly the distribution of the dark matter within these structures.

\subsubsection{HSCJ234114+031506: A potential double source plane lens} \label{subsubsec:2341}
Based on the image configuration, HSCJ234114+031506 (grade A) is a potential double source plane lens (DSPL), in which there are two distinct sources at different redshifts.  These systems are valuable due to the improved constraints on the lens model and the potential to constrain cosmological parameters (e.g., \cite{collett+2012,collett+2014,linder2016}), but are extremely rare.  Only a handful of galaxy-scale DSPLs been discovered so far, including one spectroscopically-confirmed system from the HSC SSP (the ``Eye of Horus"; \cite{tanaka+2016}).

The large northern arc in HSCJ234114+031506 is an obvious lensed feature, but it is unclear whether the compact image just south of the lens galaxy is the counterimage of this arc, or of the other nearby clump (northeast of the lens galaxy) interior to the arc.  In addition, the more distant object to the west of the lens galaxy may be a counterimage of the northeast clump.  Follow-up spectroscopy should reveal the nature of these objects.

\section{Summary} \label{sec:summary}
As part of the SuGOHI project, we have conducted a search for galaxy-scale strong gravitational lenses in DR4 of the HSC SSP (covering \ssparea~deg$^{2}$ in the $grizy$ bands) using the lens-finding algorithm {\sc YattaLens}.  We run the search on \Ngal~LRGs with spectroscopy from BOSS DR15 and visually inspect and grade the candidates identified by {\sc YattaLens} as potential lenses.  We find \NA~definite (grade A) lenses, \NB~probable (grade B) lenses, and \NC~possible (grade C) lenses.  We cross-check with recent lens searches in the literature.  The lens candidates have lens redshifts primarily in the range $0.3 \lesssim z_{\mathrm{L}} \lesssim 0.9$, consistent with previous lenses found as part of SuGOHI-g.  This intermediate redshift range makes this sample valuable for future studies of the evolution of galaxy structure from $z \sim 1$ to the present epoch.  Follow-up spectroscopy will confirm these lenses and measure source redshifts to enable science results from detailed lens modeling.

\begin{ack}
We thank the referee, whose comments helped to improve this paper.  We thank Alessandro Sonnenfeld, Sherry Suyu, and Masayuki Tanaka for useful discussions and input.  We thank Raoul Ca\~{n}ameras and Yiping Shu for providing an updated list of lens candidates from recent searches with which to cross-check our newly discovered candidates.
The Hyper Suprime-Cam (HSC) collaboration includes the astronomical communities of Japan and Taiwan, and Princeton University.  The HSC instrumentation and software were developed by the National Astronomical Observatory of Japan (NAOJ), the Kavli Institute for the Physics and Mathematics of the Universe (Kavli IPMU), the University of Tokyo, the High Energy Accelerator Research Organization (KEK), the Academia Sinica Institute for Astronomy and Astrophysics in Taiwan (ASIAA), and Princeton University.  Funding was contributed by the FIRST program from the Japanese Cabinet Office, the Ministry of Education, Culture, Sports, Science and Technology (MEXT), the Japan Society for the Promotion of Science (JSPS), Japan Science and Technology Agency (JST), the Toray Science  Foundation, NAOJ, Kavli IPMU, KEK, ASIAA, and Princeton University.
This paper is based in part on data collected at the Subaru Telescope and retrieved from the HSC data archive system, which is operated by Subaru Telescope and Astronomy Data Center (ADC) at NAOJ. Data analysis was in part carried out with the cooperation of Center for Computational Astrophysics (CfCA) at NAOJ.  We are honored and grateful for the opportunity of observing the Universe from Maunakea, which has the cultural, historical and natural significance in Hawaii.
This paper makes use of software developed for Vera C. Rubin Observatory. We thank the Rubin Observatory for making their code available as free software at http://pipelines.lsst.io/. 
Funding for SDSS-III has been provided by the Alfred P. Sloan Foundation, the Participating Institutions, the National Science Foundation, and the U.S. Department of Energy Office of Science. The SDSS-III web site is http://www.sdss3.org/.  SDSS-III is managed by the Astrophysical Research Consortium for the Participating Institutions of the SDSS-III Collaboration including the University of Arizona, the Brazilian Participation Group, Brookhaven National Laboratory, Carnegie Mellon University, University of Florida, the French Participation Group, the German Participation Group, Harvard University, the Instituto de Astrofisica de Canarias, the Michigan State/Notre Dame/JINA Participation Group, Johns Hopkins University, Lawrence Berkeley National Laboratory, Max Planck Institute for Astrophysics, Max Planck Institute for Extraterrestrial Physics, New Mexico State University, New York University, Ohio State University, Pennsylvania State University, University of Portsmouth, Princeton University, the Spanish Participation Group, University of Tokyo, University of Utah, Vanderbilt University, University of Virginia, University of Washington, and Yale University. 
This work is supported by JSPS KAKENHI Grant Numbers JP20K14511, JP20K04016, JP20H00181, JP20H05856, and JP18K03693.
J.~H.~H.~C.~acknowledges support from the Swiss National Science Foundation (SNSF).
A.~T.~J.~is supported by Riset ITB 2021.
This research made use of Astropy,\footnote{http://www.astropy.org} a community-developed core Python package for Astronomy \citep{astropy+2013,astropy+2018}.
This research made use of Matplotlib, a 2D graphics package used for Python \citep{hunter2007}.
\end{ack}

\bibliographystyle{myaasjournal}
\bibliography{sugohi8_s21a_lenses}

\appendix
\section*{Grade C Lens Candidates} \label{app:grade_c}
For completeness, we list the grade C lens candidates identified in our lens search here in Table~\ref{tab:c_candidates}.

\renewcommand*\arraystretch{1.0}
\begin{table}
\caption{Grade C Lens candidates \label{tab:c_candidates}}
\begin{minipage}{\linewidth}
\begin{tabular}{l|rrr}
\hline
Name &
$\alpha$ (J2000) &
$\delta$ (J2000) &
$z_{\mathrm{L}}$
\\
\hline
HSCJ000106+010329& 
    0.2771& 
    1.0583& 
0.7213
\\
HSCJ000158+043438& 
    0.4928& 
    4.5774& 
0.6753
\\
HSCJ000222+042125& 
    0.5957& 
    4.3571& 
0.7072
\\
HSCJ000449+041056& 
    1.2048& 
    4.1824& 
0.4934
\\
HSCJ000733+022624& 
    1.8898& 
    2.4401& 
0.4647
\\
HSCJ001144+010102& 
    2.9359& 
    1.0175& 
0.7936
\\
HSCJ001317+020931& 
    3.3243& 
    2.1588& 
0.5432
\\
HSCJ002037+002058& 
    5.1575& 
    0.3497& 
0.3288
\\
HSCJ002624+041825& 
    6.6039& 
    4.3071& 
0.7169
\\
HSCJ002652+030356& 
    6.7171& 
    3.0656& 
0.6469
\\
HSCJ003254+004839& 
    8.2257& 
    0.8109& 
0.6573
\\
HSCJ003632+014916& 
    9.1363& 
    1.8213& 
0.6477
\\
HSCJ003650+005316& 
    9.2112& 
    0.8880& 
0.6029
\\
HSCJ005356+015546& 
   13.4848& 
    1.9296& 
0.6891
\\
HSCJ005453+022522& 
   13.7214& 
    2.4230& 
0.5046
\\
HSCJ005630--001123& 
   14.1279& 
   -0.1900& 
0.6811
\\
HSCJ005937+010731& 
   14.9066& 
    1.1255& 
0.6577
\\
HSCJ010021--004349& 
   15.0899& 
   -0.7303& 
0.5548
\\
HSCJ010201+025404& 
   15.5067& 
    2.9014& 
0.5581
\\
HSCJ010434+035410& 
   16.1452& 
    3.9030& 
0.5438
\\
HSCJ012308+011027& 
   20.7850& 
    1.1743& 
0.6541
\\
HSCJ012411+025155& 
   21.0477& 
    2.8654& 
0.5084
\\
HSCJ013539+025615& 
   23.9146& 
    2.9376& 
0.5772
\\
HSCJ014101+032742& 
   25.2573& 
    3.4617& 
0.5218
\\
HSCJ020745+004721& 
   31.9391& 
    0.7894& 
0.5422
\\
HSCJ021214+002719& 
   33.0617& 
    0.4553& 
0.5372
\\
HSCJ021803--000254& 
   34.5159& 
   -0.0486& 
0.3472
\\
HSCJ022027--004453& 
   35.1139& 
   -0.7481& 
0.7743
\\
HSCJ022314--002558& 
   35.8110& 
   -0.4328& 
0.5614
\\
HSCJ022624--001716& 
   36.6001& 
   -0.2879& 
0.5540
\\
HSCJ023637+025729& 
   39.1560& 
    2.9583& 
0.6683
\\
HSCJ083759+050612& 
  129.4976& 
    5.1035& 
0.6284
\\
HSCJ083841+025720& 
  129.6733& 
    2.9556& 
0.5408
\\
HSCJ084306+051435& 
  130.7774& 
    5.2431& 
0.6232
\\
HSCJ084905+034329& 
  132.2738& 
    3.7249& 
0.4659
\\
HSCJ085247+045218& 
  133.1982& 
    4.8718& 
0.5458
\\
HSCJ093626+034716& 
  144.1122& 
    3.7878& 
0.6962
\\
HSCJ093844+015406& 
  144.6855& 
    1.9019& 
0.8345
\\
HSCJ101734--001227& 
  154.3919& 
   -0.2076& 
0.8457
\\
HSCJ102041+002226& 
  155.1714& 
    0.3739& 
0.6143
\\
\hline
\end{tabular}
\\
{\footnotesize Lens redshifts are from SDSS DR15.}
\end{minipage}
\end{table}
\renewcommand*\arraystretch{1.0}

\addtocounter{table}{-1}
\renewcommand*\arraystretch{1.0}
\begin{table}
\caption{(Continued.)}
\begin{minipage}{\linewidth}
\begin{tabular}{l|rrr}
\hline
Name &
$\alpha$ (J2000) &
$\delta$ (J2000) &
$z_{\mathrm{L}}$
\\
\hline
HSCJ102117+031442& 
  155.3225& 
    3.2450& 
0.6742
\\
HSCJ102311--000702& 
  155.7989& 
   -0.1174& 
0.4735
\\
HSCJ102858--013141& 
  157.2419& 
   -1.5282& 
0.5178
\\
HSCJ103430+042557& 
  158.6265& 
    4.4326& 
0.6546
\\
HSCJ103519+050108& 
  158.8324& 
    5.0191& 
0.6353
\\
HSCJ104146+021925& 
  160.4429& 
    2.3238& 
0.4850
\\
HSCJ104307+041425& 
  160.7815& 
    4.2403& 
0.5751
\\
HSCJ104358+022843& 
  160.9950& 
    2.4789& 
0.7851
\\
HSCJ104549--002713& 
  161.4574& 
   -0.4538& 
0.5017
\\
HSCJ104727+024906& 
  161.8667& 
    2.8184& 
0.5699
\\
HSCJ104817+004336& 
  162.0720& 
    0.7268& 
0.5811
\\
HSCJ105202+001427& 
  163.0107& 
    0.2409& 
0.6272
\\
HSCJ105329+012255& 
  163.3739& 
    1.3820& 
0.7975
\\
HSCJ105338+045536& 
  163.4101& 
    4.9267& 
0.5050
\\
HSCJ105813+031136& 
  164.5554& 
    3.1934& 
0.4728
\\
HSCJ110028+024436& 
  165.1200& 
    2.7436& 
0.5228
\\
HSCJ110114+002016& 
  165.3094& 
    0.3380& 
0.6085
\\
HSCJ110155+032954& 
  165.4827& 
    3.4985& 
0.4440
\\
HSCJ110404--003703& 
  166.0190& 
   -0.6175& 
0.5027
\\
HSCJ110511+042147& 
  166.2978& 
    4.3632& 
0.7176
\\
HSCJ110840--001927& 
  167.1686& 
   -0.3244& 
0.4505
\\
HSCJ110905+005118& 
  167.2712& 
    0.8551& 
0.6368
\\
HSCJ111230+043153& 
  168.1281& 
    4.5314& 
0.4993
\\
HSCJ111618+033628& 
  169.0772& 
    3.6079& 
0.4661
\\
HSCJ111818+021102& 
  169.5774& 
    2.1840& 
0.5180
\\
HSCJ112203--014707& 
  170.5144& 
   -1.7854& 
0.4525
\\
HSCJ112221--011759& 
  170.5890& 
   -1.3000& 
0.7005
\\
HSCJ112710+025956& 
  171.7928& 
    2.9991& 
0.5387
\\
HSCJ113838+050110& 
  174.6612& 
    5.0196& 
0.6475
\\
HSCJ114358+022513& 
  175.9927& 
    2.4205& 
0.6342
\\
HSCJ114847+032706& 
  177.1993& 
    3.4518& 
0.6271
\\
HSCJ115247+034254& 
  178.1973& 
    3.7152& 
0.5721
\\
HSCJ115313+030814& 
  178.3060& 
    3.1373& 
0.5844
\\
HSCJ115446+042415& 
  178.6934& 
    4.4042& 
0.7196
\\
HSCJ115907+051911& 
  179.7829& 
    5.3200& 
0.4692
\\
HSCJ120233+042541& 
  180.6395& 
    4.4281& 
0.5077
\\
HSCJ120255+034044& 
  180.7314& 
    3.6791& 
0.5256
\\
HSCJ120745+041400& 
  181.9380& 
    4.2334& 
0.6194
\\
HSCJ121040+033354& 
  182.6697& 
    3.5652& 
0.7540
\\
HSCJ122843+023818& 
  187.1824& 
    2.6385& 
0.5260
\\
\hline
\end{tabular}
\\
\end{minipage}
\end{table}
\renewcommand*\arraystretch{1.0}

\addtocounter{table}{-1}
\renewcommand*\arraystretch{1.0}
\begin{table}
\caption{(Continued.)}
\begin{minipage}{\linewidth}
\begin{tabular}{l|rrr}
\hline
Name &
$\alpha$ (J2000) &
$\delta$ (J2000) &
$z_{\mathrm{L}}$
\\
\hline
HSCJ122948+013652& 
  187.4527& 
    1.6146& 
0.6183
\\
HSCJ123035+052251& 
  187.6474& 
    5.3810& 
0.4701
\\
HSCJ124503+024354& 
  191.2632& 
    2.7319& 
0.4949
\\
HSCJ124629+004116& 
  191.6211& 
    0.6880& 
0.5730
\\
HSCJ125220+025504& 
  193.0858& 
    2.9180& 
0.5178
\\
HSCJ125604--003229& 
  194.0177& 
   -0.5415& 
0.7249
\\
HSCJ125903+014542& 
  194.7630& 
    1.7618& 
0.5178
\\
HSCJ130044+031657& 
  195.1864& 
    3.2826& 
0.6798
\\
HSCJ130500+032545& 
  196.2540& 
    3.4293& 
0.5266
\\
HSCJ131037+021947& 
  197.6551& 
    2.3300& 
0.5056
\\
HSCJ131140+025539& 
  197.9179& 
    2.9276& 
0.6115
\\
HSCJ131417--010555& 
  198.5711& 
   -1.0989& 
0.4833
\\
HSCJ131519+040008& 
  198.8315& 
    4.0024& 
0.7557
\\
HSCJ131604+040924& 
  199.0194& 
    4.1569& 
0.5786
\\
HSCJ131626+020910& 
  199.1089& 
    2.1530& 
0.6845
\\
HSCJ131759+025548& 
  199.4975& 
    2.9302& 
0.6018
\\
HSCJ131933--004256& 
  199.8885& 
   -0.7158& 
0.6575
\\
HSCJ132221--012821& 
  200.5878& 
   -1.4727& 
0.5423
\\
HSCJ132439--015026& 
  201.1664& 
   -1.8407& 
0.4379
\\
HSCJ132904+020015& 
  202.2679& 
    2.0044& 
0.5195
\\
HSCJ133328+425215& 
  203.3694& 
   42.8711& 
0.5873
\\
HSCJ134017+420751& 
  205.0742& 
   42.1311& 
0.5370
\\
HSCJ134504--012202& 
  206.2705& 
   -1.3673& 
0.5555
\\
HSCJ135110+443315& 
  207.7926& 
   44.5544& 
0.5650
\\
HSCJ135346+021338& 
  208.4447& 
    2.2274& 
0.4589
\\
HSCJ140218+022318& 
  210.5786& 
    2.3884& 
0.6941
\\
HSCJ140252+023101& 
  210.7177& 
    2.5171& 
0.6017
\\
HSCJ140347+425700& 
  210.9459& 
   42.9503& 
0.3946
\\
HSCJ141628+030004& 
  214.1204& 
    3.0013& 
0.4899
\\
HSCJ141758+022347& 
  214.4942& 
    2.3965& 
0.5219
\\
HSCJ142240+441140& 
  215.6670& 
   44.1946& 
0.5859
\\
HSCJ142532+032850& 
  216.3843& 
    3.4806& 
0.5294
\\
HSCJ142615+033050& 
  216.5661& 
    3.5139& 
0.5237
\\
HSCJ142652+433113& 
  216.7175& 
   43.5204& 
0.5770
\\
HSCJ142820+031902& 
  217.0869& 
    3.3175& 
0.5933
\\
HSCJ143018+034441& 
  217.5776& 
    3.7448& 
0.6769
\\
HSCJ144411+030655& 
  221.0478& 
    3.1153& 
0.7677
\\
HSCJ144530+434129& 
  221.3752& 
   43.6915& 
0.5332
\\
HSCJ145130+042054& 
  222.8785& 
    4.3484& 
0.7004
\\
HSCJ220958+013722& 
  332.4942& 
    1.6230& 
0.6250
\\
\hline
\end{tabular}
\\
\end{minipage}
\end{table}
\renewcommand*\arraystretch{1.0}

\addtocounter{table}{-1}
\renewcommand*\arraystretch{1.0}
\begin{table}
\caption{(Continued.)}
\begin{minipage}{\linewidth}
\begin{tabular}{l|rrr}
\hline
Name &
$\alpha$ (J2000) &
$\delta$ (J2000) &
$z_{\mathrm{L}}$
\\
\hline
HSCJ221655+035733& 
  334.2325& 
    3.9594& 
0.5469
\\
HSCJ222653+035502& 
  336.7215& 
    3.9173& 
0.6978
\\
HSCJ222951--012440& 
  337.4645& 
   -1.4111& 
0.7065
\\
HSCJ223315+041108& 
  338.3137& 
    4.1858& 
0.6318
\\
HSCJ223513+034313& 
  338.8063& 
    3.7204& 
0.6316
\\
HSCJ224101+041820& 
  340.2567& 
    4.3056& 
0.5916
\\
HSCJ231207+043501& 
  348.0295& 
    4.5839& 
0.5597
\\
HSCJ231740+040317& 
  349.4181& 
    4.0548& 
0.5851
\\
HSCJ232546--003916& 
  351.4424& 
   -0.6547& 
0.6010
\\
HSCJ232833+024214& 
  352.1385& 
    2.7040& 
0.4926
\\
HSCJ233219+022838& 
  353.0827& 
    2.4772& 
0.6517
\\
HSCJ233917+010908& 
  354.8243& 
    1.1523& 
0.6262
\\
HSCJ233945+015726& 
  354.9379& 
    1.9575& 
0.5046
\\
HSCJ234246+044345& 
  355.6932& 
    4.7292& 
0.4915
\\
HSCJ234824+025357& 
  357.1015& 
    2.8993& 
0.5498
\\
HSCJ235052+035356& 
  357.7167& 
    3.8990& 
0.5145
\\
HSCJ235503+014443& 
  358.7653& 
    1.7453& 
0.5224
\\
HSCJ235730+010133& 
  359.3776& 
    1.0260& 
0.6379
\\
\hline
\end{tabular}
\\
\end{minipage}
\end{table}
\renewcommand*\arraystretch{1.0}

\end{document}